THE MOTIONS OF CLUSTERS AND GROUPS OF GALAXIES

Neta A. Bahcall,[1] Mirt Gramann,[1,2] and Renyue Cen[1]

email: cen@astro.princeton.edu

[1] Princeton University Observatory, Princeton, NJ 08544
[2] Tartu Astrophysical Observatory, EE-2444 Tõravere, Estonia

## ABSTRACT

The distributions of peculiar velocities of rich clusters and of groups of galaxies are investigated for different cosmological models and are compared with observations. Four cosmological models are studied: standard ($\Omega = 1$) CDM, low-density ($\Omega = 0.3$) CDM, HDM ($\Omega = 1$), and PBI ($\Omega = 0.3$). All models are normalized to the microwave background fluctuations observed by COBE. We find that rich clusters of galaxies exhibit a Maxwellian distribution of peculiar velocities in all models, as expected from a Gaussian initial density field. The clusters appear to be efficient tracers of the large-scale velocity-field. The cluster 3-D velocity distribution typically peaks at $v \sim 600$ km s$^{-1}$, and extends to high cluster velocities of $v \sim 2000$ km s$^{-1}$. The low-density CDM model exhibits somewhat lower velocities: it peaks at $\sim 400$ km s$^{-1}$, and extends to $\sim 1200$ km s$^{-1}$. Approximately 10% ($\sim$1% for low-density CDM) of all model rich clusters move with high peculiar velocities of $v \geq 10^3$ km s$^{-1}$. The highest velocity clusters frequently originate in dense superclusters. The model velocity distributions of rich clusters are compared with the model velocity distributions of small groups of galaxies, and of the total matter. The group velocity distribution is, in general, similar to the velocity distribution of the rich clusters. The matter velocity distribution is similar to that of the rich clusters for the $\Omega = 0.3$ models; these models exhibit Maxwellian velocity distributions for clusters, for groups, and for matter that are all similar to each other. The mass distribution in the $\Omega = 1$ models, however, exhibits a longer tail of high velocities than do the clusters. This high-velocity tail originates mostly from the high velocities that exist *within* rich clusters.

The model velocity distributions of groups and clusters of galaxies are compared with observations. The data exhibit a larger high-velocity tail, to $v_r \geq 2000$ km s$^{-1}$, than seen in the model simulations (except HDM). Due to the large observational uncertainties, however, the data are consistent at a $\sim 1$ to $3\sigma$ level with the model predictions and with a Gaussian initial density field. Accurate observations of cluster peculiar velocities, especially at the high-velocity tail, should provide powerful constraints on the cosmological models.

## 1. INTRODUCTION

Clusters of galaxies are an efficient tracer of the large-scale structure of the universe. Measurements of the strong correlation function of clusters of galaxies (Bahcall & Soneira 1983; Klypin & Kopylov 1983; Postman *et al.* 1992; Peacock & West 1992; Bahcall & West 1992; Dalton *et al.* 1992; Nichol *et al.* 1992), and the superclustering properties of clusters (Bahcall & Soneira 1984, Postman *et al.* 1992, Einasto *et al.* 1994, Rhoads *et al.* 1994) provided some of the first evidence for the existence of organized structure on large scales. The existence of large-scale structure produces gravitationally-induced large-scale peculiar velocities of galaxies, of groups of galaxies, and of clusters of galaxies. Large-scale peculiar motions of galaxies have been detected (Rubin *et al.* 1976, Dressler *et al.* 1987, Burstein *et al.* 1987, Faber *et al.* 1989); combined with observations of the large-scale distribution of galaxies, these peculiar velocities provide important constraints on cosmological models (Bertschinger & Dekel 1989; Kofman *et al.* 1993; Croft & Efstathiou 1993; Dekel 1994 and references therein).

The motions of clusters of galaxies have also been investigated. Bahcall *et al.* (1986) suggested the possible existence of large peculiar velocities of clusters in some dense superclusters, with $v_r \sim 10^3$ km s$^{-1}$. Faber *et al.* (1989) and Mould *et al.* (1991, 1993) find, using direct distance indicators, clusters with similarly high peculiar velocities, $v_r \sim 10^3$ km s$^{-1}$, in some regions. Lauer and Postman (1994) have recently reported a large bulk-flow of relatively nearby rich clusters of galaxies, with $v_r \sim 700$ km s$^{-1}$.

What is the role of rich clusters of galaxies in mapping the velocity field of the universe? Do rich clusters provide a useful tracer of the large-scale velocity field? In this paper, we study the velocity distribution of rich clusters of galaxies in several popular cosmological models. We also compare the velocity distribution of rich clusters with that of groups and of the total matter for each of the models. We conclude that rich clusters of galaxies are efficient tracers of large-scale velocities. We compare the model expectations with observations of group and cluster velocities.

We describe the model simulations in §2. In §3 we present the distribution of peculiar velocities of rich clusters of galaxies in the models. We compare in §4 the rich-cluster velocity distribution with that of small groups of galaxies and of the total matter distribution. We present in §5 a fit of the cluster velocity distribution to a Maxwellian. In §6 we investigate the velocity distribution of clusters within dense superclusters in the models. We compare the cluster and group velocity distribution with observations in §7, and summarize the main conclusions in §8.

## 2. MODEL SIMULATIONS

A large-scale Particle-Mesh code with a box size of 800h$^{-1}$ Mpc is used to simulate the evolution of the dark matter in different cosmological models. A large simulation box is needed in order to ensure that contributions to velocities from waves larger than the box size are small, and to minimize uncertainties due to fluctuations in the small number of large waves (see below). The simulation box contains $500^3$ cells and $250^3 = 10^{7.2}$ dark matter particles. The spatial resolution is 1.6h$^{-1}$ Mpc. [A higher resolution (0.8h$^{-1}$ Mpc) smaller box (400h$^{-1}$ Mpc) is also studied for comparison.] Details of the simulations are discussed in Cen (1992).

Four cosmological model simulations are performed: Standard Cold Dark-Matter (CDM) model ($\Omega = 1$); low-density CDM model ($\Omega = 0.3$); Hot Dark-Matter model (HDM, $\Omega = 1$); and Primeval Baryonic Isocurvature model (PBI, $\Omega = 0.3$). The specific model parameters are summarized in Table 1. These parameters include the matter density, $\Omega$; the cosmological constant contribution, $\Omega_\Lambda$; the Hubble constant (in units of $H_O = 100$h km s$^{-1}$ Mpc$^{-1}$); and the normalization of the mass fluctuations on 8h$^{-1}$ Mpc scale, $\sigma_8$. The models are normalized to the large-scale microwave background anisotropy measured by COBE (Smoot *et al.* 1992). (The HDM model normalization is $\sim 20\%$ higher than the $\Omega = 1$ CDM on large scales). The parameters of the PBI and low-density CDM models are described in detail by Cen, Ostriker, & Peebles (1993) and Cen, Gnedin, & Ostriker (1993). Both the CDM and HDM models use the adiabatic density perturbation power spectra given by Bardeen *et al.* (1986).

Clusters are selected in each simulation using an adaptive linkage algorithm described by Suto, Cen & Ostriker (1992) and Bahcall & Cen (1992). The cluster mass threshold, within $r = 1.5\mathrm{h}^{-1}$ Mpc of the cluster center, is selected to correspond to a number density of clusters comparable to the observed density of rich ($R \geq 1$) clusters, $n_{cl} \sim 6 \times 10^{-6}$ h$^3$ Mpc$^{-3}$ (Bahcall & Cen 1993). When analyzing smaller groups of galaxies (§4), the threshold is appropriately reduced to correspond to the observed number density of small groups of galaxies, $n_{gr} \sim 10^{-4}$ h$^3$ Mpc$^{-3}$ (Ramella *et al.* 1989; Bahcall & Cen 1993).

A total of $\sim 3000$ rich clusters of galaxies and $\sim 5 \times 10^4$ groups are obtained in each of the $800\mathrm{h}^{-1}$ Mpc simulation models. The three-dimensional peculiar velocity of each of these clusters (or groups), relative to the co-moving cosmic background frame, is determined from the simulation; these velocities are used in the analyses described below. The analysis is carried out independently for rich clusters (§3), groups (§4), and matter (§4), in order to compare the velocity distributions of the different systems.

The simulation results are consistent with expectations from linear theory (see §4). Using linear theory we find that contributions from waves larger than $800\mathrm{h}^{-1}$ Mpc are small; the rms velocities integrated to $\lambda = 800\mathrm{h}^{-1}$ Mpc and to infinity are similar to each other within $\leq 10$ km s$^{-1}$ Mpc. We also study the large-scale bulk velocities in the simulations and in linear theory. We find that the simulated and expected bulk-flows are consistent with each other to within $\leq 60$ km s$^{-1}$ on $50\mathrm{h}^{-1}$ Mpc radius scale (Gramann *et al.* 1994; Strauss *et al.* 1994). The sensitivity of the results to the resolution of the simulation is tested by comparing simulations with $0.8\mathrm{h}^{-1}$ Mpc nominal resolution with those of $1.6\mathrm{h}^{-1}$ Mpc nominal resolution (both in a $400\mathrm{h}^{-1}$ Mpc box). The resulting group and cluster velocity distributions are consistent in both resolutions to within $\leq 60$ km s$^{-1}$.

## 3. PECULIAR VELOCITIES OF CLUSTERS OF GALAXIES

How fast do rich clusters of galaxies move? We present in Figure 1 the results for the integrated peculiar velocity distribution of rich clusters for each of the models studied. The integrated velocity distribution represents the probability distribution, or the normalized number density, of clusters with peculiar velocities larger than $v$, $P(> v)$. The velocity $v$ refers to the three-dimensional peculiar velocity of the cluster relative to the cosmic background frame.

Two results are immediately apparent from Figure 1. First, the cluster velocity distribution, which exhibits a moderate fall-off at high velocities ($v \sim 500 - 2000$ km s$^{-1}$), and a leveling-off at small velocities ($v \leq 500$ km s$^{-1}$), is similar in all models except $\Omega = 0.3$ CDM; the latter exhibits lower velocities than the other models. Second, a significant fraction of all model rich clusters ($\sim 10\%$) exhibit high peculiar velocities of $v \geq 10^3$ km s$^{-1}$ ($v \geq 700$ km s$^{-1}$ for $\Omega = 0.3$ CDM). The tail of the cluster velocity distribution reaches $\sim 2000$ km s$^{-1}$ ($\sim 1200$ km s$^{-1}$ for $\Omega = 0.3$ CDM).

The differential velocity distribution of the clusters, $P(v)$ (i.e., the normalized number density of clusters with peculiar velocity in the range $v \pm dv$, per unit $dv$, as a function of $v$), is presented in Figure 2. The cluster velocity distribution is similar for all but the low-density CDM model, as is expected from Figure 1. The specific shape of the velocity distribution is also of interest. The cluster velocities peak at $v \sim 600$ km s$^{-1}$ ($\sim 400$ km s$^{-1}$ for $\Omega = 0.3$ CDM). The distribution decreases rapidly at lower

Figure 1. Integrated 3-D peculiar velocity distribution of rich clusters of galaxies in four cosmological models: $\Omega = 1$ CDM (solid line), $\Omega = 0.3$ CDM (dashed line), $\Omega = 1$ HDM (dash-dot line), and $\Omega = 0.3$ PBI (dotted line).

velocities, and decreases more slowly at higher velocities, with a high velocity tail to $\sim 1000 - 2000$ km s$^{-1}$. The rms peculiar velocities of clusters in the different models are listed in the first column of Table 2. The rms velocities are typically in the range $\sim 700 - 800$ km $s^{-1}$ for the $\Omega = 1$ CDM, HDM, and PIB models (the HDM spectrum normalization is $\sim 20\%$ higher than that of the standard CDM on large scales, thus

Figure 2. Differential 3-D peculiar velocity distribution of rich clusters of galaxies for the four cosmological models of Fig. 1 (same notation).

yielding somewhat larger velocities); and 440 km $s^{-1}$ for the $\Omega = 0.3$ CDM model.

What causes the cluster velocity distribution to peak at a few hundred km s$^{-1}$ and exhibit the shape seen above? What is the origin of the large peculiar velocities of clusters, $v \geq 10^3$ km s$^{-1}$? We address these questions in §5 and §6.

## 4. PECULIAR VELOCITIES OF CLUSTERS, GROUPS, AND MATTER

In this section we compare the velocity distribution of rich clusters of galaxies, discussed above, with the velocity distributions of poorer systems, such as poor clusters and small groups of galaxies, as well as with the velocity of the underlying matter distribution.

Groups of galaxies are selected as described in §2, using an adaptive linkage algorithm, and a mass threshold corresponding to a total number density of $n_{gr} \sim 10^{-4}$ groups $\mathrm{Mpc}^{-3}$. About $\sim 5 \times 10^4$ such groups are identified in each of the simulations. The velocity distribution function for groups is then determined as described in §3.

The velocity distribution of all the matter (i.e., the velocities of all the dark-matter particles) in each simulation is also determined. The matter velocity distribution may represent the closest approximation to the velocity distribution of galaxies.

The differential and integrated velocity distributions of the groups and of all the matter in each of the four models are presented in Figures 3(a-d) and 4(a-d). In each sub-figure (a to d), the velocity distribution of the rich clusters, groups, and matter are compared with each other for a given model.

The differential velocity distribution of the clusters, groups, and matter (Figure 3) reveals - for all models - a peaked distribution. The distributions typically peak at peculiar velocities around 600 km $\mathrm{s}^{-1}$ (lower for $\Omega = 0.3$ CDM), and exhibit an extended tail to high velocities of $\geq 10^3$ km $\mathrm{s}^{-1}$. A comparison of the velocity distributions in the different models yields interesting results. First, in the low density models ($\Omega = 0.3$ CDM and PBI), the velocity distributions of the clusters, groups, and matter are all similar to each other. This is clearly seen in the differential (Figures 3b, 3d) and in the integrated (4b, 4d) velocity distributions. (The matter distribution is marginally shifted to higher velocities than the groups and clusters). The $\Omega = 1$ models (CDM, HDM) reveal a significant difference between the rich clusters and the matter velocity distributions; the matter exhibits a stronger tail to large velocities than do the rich clusters (Figures 3a, 3c; 4a, 4c). The effect is seen most clearly in the integrated function, Figure 4. The peak of the matter velocity distribution ($\Omega = 1$ models) is also shifted to a somewhat higher velocity ($\sim$ 800 km $\mathrm{s}^{-1}$) than the rich clusters ($\sim$ 600 km $\mathrm{s}^{-1}$). The velocity distributions of small groups are typically intermediate between the velocity distributions of the clusters and of the matter. The rms peculiar velocities of clusters, groups, and matter are summarized for all the models in Table 2.

What is the origin of the high-velocity tail in the velocity distribution of matter in the $\Omega = 1$ models? The matter (or galaxy) velocity distribution includes the velocities of the matter particles (or galaxies) *within* rich clusters. The 3-D velocity dispersion within rich clusters is very high for $\Omega = 1$ models such as CDM (with $\sigma_8 = 1$), reaching $\sim$ 3000 km $\mathrm{s}^{-1}$ (e.g., Bahcall & Cen 1992). These large velocities provide the main contribution to the high velocity tails seen in the $\Omega = 1$ models (as well as the excess seen in the matter velocity of the PBI model). Other than this additional small-scale mass contribution, the galaxies and clusters appear to trace the same velocity distribution on large scales. This is tested by smoothing the matter distribution with a Gaussian of radius $R_G = 3\mathrm{h}^{-1}$ Mpc. In Figure 5 we present the peculiar velocity distribution of clusters, of matter, and of the $3\mathrm{h}^{-1}$ Mpc smoothed matter for the CDM models ($\Omega = 1$ and $\Omega = 0.3$). The results indicate that the velocity tail of the matter distribution is eliminated by the smoothing; the smoothed matter and the clusters exhibit the same velocity distribution. The rms peculiar velocities of the smoothed mass distributions are presented for all models in Table 2; they are consistent with the cluster peculiar velocities (to within 7 to 64 km $s^{-1}$) in all models.

Figure 3. Differential velocity distribution (3-D) of rich clusters (solid line), of groups (dashed line), and of the total matter (dotted line), for different models: (a) $\Omega = 1$ CDM; (b) $\Omega = 0.3$ CDM; (c) $\Omega = 1$ HDM; and (d) $\Omega = 0.3$ PBI.

Figure 4. Same as Figure 3 but for the integrated velocity distribution (of rich clusters, groups, and matter).

The low-density CDM and PBI models do not show a high velocity tail for the matter distribution. The models velocity distributions for groups, for rich clusters, and for all the matter are similar (Figures 3, 4). In these models, the internal velocity-dispersions in clusters must therefore be similar to (or smaller than) the high velocity tail of the clusters themselves. This is indeed so; the 3-D velocity dispersion within rich clusters in the $\Omega = 0.3$ CDM model, $\leq 1500$ km s$^{-1}$ (consistent with cluster observations; Bahcall & Cen 1992), is comparable to the tail of the cluster peculiar velocities (Figures 3b, 4b). (See also Figure 5 and Table 2).

Figure 5. Differential 3-D peculiar velocity distribution of rich clusters (solid line), of the total matter (dotted line), and of the matter smoothed with a Gaussian radius $R_G = 3h^{-1}$ Mpc (dash - dot line). (a) $\Omega = 1$ CDM; (b) $\Omega = 0.3$ CDM.

The simulation results were tested against expectations from linear theory calculations. We compare the rms peculiar velocities of the smoothed $R_G = 3h^{-1}$ Mpc matter distribution in each model with the velocities expected from linear theory when smoothed by the same smoothing length. The results are listed in Table 2. The linear theory values compare well with the simulated smoothed matter velocities for all models (typically within $\leq 50$ km s$^{-1}$).

## 5. MAXWELLIAN FITS OF THE VELOCITY DISTRIBUTION

The velocity distributions of clusters, of groups, and of matter are fitted with Maxwellian distributions $P(v) \propto v^2 \exp(-v^2/2\sigma^2)$. The velocity dispersion for each case ($\sigma$) is determined from the model simulation. The results are presented in Figures 6 and 7. The velocity distributions of clusters and of all the matter (dark solid and dotted lines, respectively) are compared with the Maxwellian distribution (faint lines). The differential (Figure 6) and the integrated (Figure 7) velocity distributions are shown for the $\Omega = 1$ and $\Omega = 0.3$ CDM models.

The velocity distribution of clusters of galaxies is well represented by a Maxwellian for *all* models (CDM, HDM, and PBI). Small deviations from a Maxwellian are suggested at the highest velocities; these are mostly due to clusters that are located in dense superclusters (see §6). The velocity distribution of the matter fits a Maxwellian only for the low-density CDM model (where it is similar to the velocity distribution of the clusters; Figures 6b, 7b). The matter velocity distribution in the $\Omega = 1$ models can not be represented by a Maxwellian - it exhibits a more extended tail at high velocities (Figures 6a, 7a). This tail is contributed mostly by the high velocity-dispersion matter within rich clusters in this model (§4); it is best approximated by an exponential fit (Cen & Ostriker 1993).

The HDM model yields similar results to those of the $\Omega = 1$ CDM: the clusters are well fit by a Maxwellian distribution, but the matter velocities exhibit a more extended tail than expected in a Maxwellian distribution. The PBI model resembles more closely the $\Omega = 0.3$ CDM results.

A Maxwellian distribution of the large-scale velocity field is expected from Gaussian initial density fluctuations. The above results suggest that clusters of galaxies, with their Maxwellian velocity distribution, provide a fundamental tracer of the large-scale velocity field. The cluster velocities can be used to test the type of the initial density field (Gaussian or non-Gaussian).

## 6. PECULIAR VELOCITIES OF CLUSTERS IN SUPERCLUSTERS

Approximately 10% of all rich clusters of galaxies in the models exhibit large peculiar velocities, $v \geq 10^3$ km s$^{-1}$ (or $\geq 700$ km s$^{-1}$ for $\Omega = 0.3$ CDM) (§3). What is the origin of these large velocities? Do they originate from the gravitational interaction in dense superclusters (or close cluster pairs) as suggested by Bahcall, Soneira & Burgett (1986)? We investigate this question below.

We identify superclusters (groups of clusters of galaxies) in each of the model simulations using a linkage algorithm for the rich clusters (§2,3). A linkage-length of 10h$^{-1}$ Mpc is used; it identifies dense superclusters. All rich cluster pairs separated by less than 10h$^{-1}$ Mpc are grouped into a "supercluster". The superclusters contain, by definition, two or more rich clusters. This procedure follows the standard observational selection of superclusters (e.g., Bahcall and Soneira 1984).

The velocity distribution of clusters that are supercluster members is determined for each model. This distribution is compared with the velocity distribution of *all* clusters (i.e., not just those that are supercluster members), as well as of the *isolated* clusters (i.e., *not* in superclusters).

Figure 6. Differential velocity distribution (3-D) of clusters and of the total matter in the models (dark solid and dotted lines, respectively), and their comparison with a Maxwellian distribution (faint lines) (§5). (a) $\Omega = 1$ CDM; (b) $\Omega = 0.3$ CDM.

The results are presented in Figure 8. The velocity distributions of isolated clusters, clusters in superclusters, and all clusters are compared with each other. Figures 8a-b present the integrated velocity distribution for the $\Omega = 1$ and $\Omega = 0.3$ CDM models. The results show, as expected, that the velocity distribution of clusters in dense superclusters differs somewhat from that of isolated clusters. For $\Omega = 1$ CDM, only $\sim 10\%$ of all isolated clusters have velocities $\geq 10^3$ km s$^{-1}$; however, $\sim 50\%$ of supercluster members exhibit these same high velocities. The effect is similar, though somewhat smaller in magnitude, in the low-density CDM model: $\sim$ 12% of isolated clusters versus $\sim$ 30% of supercluster members have velocities $\geq$ 600 km s$^{-1}$.

Figure 7. Same as Figure 6 but for the integrated velocity distribution.

## 7. COMPARISON WITH OBSERVATIONS

Observational determination of peculiar velocities of galaxies, groups, and clusters is difficult, since the true distances of the objects and hence their Hubble velocities are uncertain. However, data are available for the peculiar velocities of some samples of groups and of clusters of galaxies. We use the group and cluster peculiar velocities measured by Aaronson *et al.* (1986), Mould *et al.* (1991, 1993), and Mathewson *et al.*

Figure 8. Integrated velocity distribution (3-D) of model clusters that are: members of dense superclusters (dashed line); isolated clusters (dotted line); and all clusters (solid line) (§6). (a) $\Omega = 1$ CDM; (b) $\Omega = 0.3$ CDM.

(1992), who employ the Tully-Fisher (TF) method for distance indicators, and Faber *et al.* (1989), who use the $D_n - \sigma$ method. Groups with observational velocity uncertainties $\geq 900$ km s$^{-1}$ are excluded from the analysis. A total of 48 group and cluster peculiar velocities are available from the TF method, and 91 from $D_n - \sigma$. The total sample of peculiar velocities includes 123 non-overlapping groups and clusters. These data are used to determine the observed velocity distribution of groups of galaxies. The data correspond to low-threshold groups, with populations comparable to the simulated groups studied above. As seen in §3-4, the velocity distribution is insensitive to the group threshold: small groups and rich clusters generally yield similar results.

The observed differential and integrated group velocity distributions are presented in Figures 9 to 11. Here we use the one-dimensional velocities, as observed ($v_{1D}$); they are compared with the 1-D velocities in the models (as opposed to the 3-D velocities used in the previous sections, since the actual 3-D velocities can not be directly observed). We present in different symbols the data obtained from the TF, $D_n - \sigma$, and total (combined) samples. The results from the different sub-samples are approximately consistent with each other (with $D_n - \sigma$ exhibiting a somewhat higher velocity tail than TF at the highest velocities).

The number of rich clusters in the observed sample is rather small ($\sim 18$ $R \geq 0$ clusters). Within the large statistical uncertainties of such a small sample, the observed rich cluster velocity distribution is consistent with that of the groups, and thus consistent with the model comparisons discussed below.

The observed velocity distribution is superimposed on the 1-D velocity distribution expected for groups for each of the four models (dotted line; Figs. 9a-d to 11a-d). The shape of the functions differs from those of Figures 3-5 since the 1-D velocity distribution is plotted instead of the 3-D. The 3-D distribution is proportional to $v^2$ at small velocities; instead, the 1-D velocities exhibit a Gaussian distribution, as expected for a 3-D Maxwellian. In comparing model expectations with observations, the model velocity distribution (dotted line) needs to be convolved with the observational uncertainties; each model cluster is thus given an uncertainty drawn at random from the *actual* distribution of observed uncertainties of the total sample. The convolved model distribution of groups is shown by the dashed lines in Figs. 9-11; the convolved rich cluster distributions are shown by the solid lines, for comparison. The convolution flattens the model distributions, as expected, and produces a high-velocity tail. The convolution also reduces the differences between the different model distributions, as well as between groups and clusters. The rms peculiar velocities of model groups are compared with observations (total sample and T-F alone) in Table 3. The models are convolved with the observational uncertainties of each sample (total and TF) separately.

A comparison between the data and the convolved models suggests that the observed and model velocity distributions are consistent with each other at $\sim 1$ to $3\sigma$ level. A K-S test of the velocity distribution indicates that the models are consistent with the data at significance levels of $\sim 5\%$, 3%, 15% and 1% for $\Omega = 1$ CDM, $\Omega = 0.3$ CDM, PBI and HDM, respectively. The significance levels rise to $\sim 46\%$, 70%, 83%, and 7% for the same models if only the T-F data are used.

The observations (especially the $D_n - \sigma$ data) exhibit a long tail of high velocity groups and clusters, to $v_{1D} \geq 2000$ km s$^{-1}$. This high-velocity tail is inconsistent with all but the HDM model. The T-F data, however, exhibits a lower velocity tail than the $D_n - \sigma$ counterpart; it is more consistent with the convolved models. Since the current observational uncertainties are large, and the effect of model convolution is strong (in fact it yields most of the high-velocity tail), more accurate velocity data are needed in order to further constrain the cosmological models. One effect of large observational errors is to produce an artificial high-velocity tail. It therefore seems likely that more accurate cluster velocities will yield *smaller velocities than currently suggested by the observed high-velocity tail* of Figures 10-11. If, however, the high velocities ($v_{1D} > 2000$ km $s^{-1}$) are confirmed, with $P > 0.01$, it will suggest that some of the above models (e.g., low-density CDM) are unlikely.

Figure 9. Comparison of observations and model simulations. The observed differential velocity distribution (in 1-D velocities) of groups of galaxies as determined from Tully-Fisher (TF) and $D_n - \sigma$ distance-indicators (stars and open circles, respectively), and for the combined sample (dark circles with $\sqrt{N}$ uncertainties indicated) are presented (§7). The model 1-D group velocity distribution (dotted line), and its convolved distribution (dashed line; convolved with the observational velocity uncertainties of the total sample) are shown. The convolved distribution of simulated rich clusters (solid line) is also shown, for comparison. The observations should be compared with the convolved model group simulations (dashed line). (a) $\Omega = 1$ CDM, (b) $\Omega = 0.3$ CDM; (c) $\Omega = 1$ HDM; (d) $\Omega = 0.3$ PBI.

Figure 10. Same as Figure 9, but in a logarithmic scale.

The $\Omega = 1$ CDM model has known problems: its high COBE normalization used here ($\sigma_8 = 1.05$) is inconsistent with the mass-function and correlation-function of clusters of galaxies, the angular correlation function of galaxies, and with the small-scale pair-wise galaxy velocities (Maddox *et al.* 1989; Bahcall & Cen 1992; Kofman *et*

Figure 11. Same as Figure 10, but for the integrated velocity distributions. (The $\sqrt{N}/N$ of each bin is presented, for illustration only, by the vertical bars).

*al.* 1993; Ostriker 1993). A lower normalization will yield a lower velocity tail than given above (and will still be inconsistent with the cluster correlations and mass-function). The COBE normalized HDM model has the problem of late galaxy formation. An $\Omega = 1$ mixed HDM+CDM model (30% + 70%, respectively), properly normalized to COBE, will yield a velocity distribution comparable to that of the $\Omega \sim 1$ CDM model.

The COBE-normalized low-density CDM model is consistent with many current observations of large-scale structure, including the galaxy and the cluster correlation functions, the power spectrum of galaxies, the small-scale peculiar velocities of galaxies, and the mass-function of galaxy clusters (Maddox *et al.* 1989; Bahcall & Cen 1992; Efstathiou *et al.* 1992, Kofman *et al.* 1993; Ostriker 1993). This model is also consistent (at a K-S significance level of $\sim 70\%$ for the T-F data, or $\sim 3\%$ for the total sample) with the observed peculiar velocity distribution of groups and clusters studied above. Accurate determination of the high peculiar velocity tail of the cluster distribution will provide a critical test of this model.

## 8. CONCLUSIONS

The distributions of peculiar velocities of rich clusters of galaxies, of groups, and of the total matter have been investigated using large-scale simulations of four cosmological models: standard CDM ($\Omega = 1$), low-density CDM, HDM ($\Omega = 1$), and PBI. The main conclusions are summarized below.

(1) Rich clusters of galaxies exhibit a robust, Maxwellian distribution of peculiar velocities in all models studied. The distribution typically peaks at $v \sim 600$ km s$^{-1}$, and extends to high velocities of $v \sim 2000$ km s$^{-1}$. The velocity distribution is similar in all models except low-density CDM, which is shifted to lower velocities. Approximately 10% of all rich clusters move with peculiar velocities of $v \geq 10^3$ km s$^{-1}$ ($\geq 700$ km s$^{-1}$ for $\Omega = 0.3$ CDM).

(2) The highest velocity clusters, with $v > 10^3$ km s$^{-1}$, originate frequently in dense superclusters.

(3) The velocity distribution of model clusters is insensitive to the cluster selection threshold (i.e., richness). The velocity distribution of small groups of galaxies is similar to that of the rich clusters, with only a minor suggested shift to larger velocities (larger shift for HDM).

(4) The velocity distribution of the total matter is similar to the velocity distribution of groups and of rich clusters for the low-density models. In the $\Omega = 1$ models the total mass exhibits a larger tail of high-velocities ($\geq 2000 - 3000$ km s$^{-1}$ ) than do the clusters. This high-velocity tail of the mass distribution reflects the large velocity-dispersions that exist within rich clusters of galaxies in high-density models. The mass velocity distribution for the $\Omega = 1$ models can not be described by a Maxwellian; it is better approximated by an exponential. The low-density models, on the other hand, exhibit velocity distributions of clusters, of groups and of the total matter that are all similar to each other, and that are each described well by a Maxwellian distribution.

(5) The observed distributions of peculiar velocities of groups and of clusters of galaxies are consistent at $\sim 1$ to $3\sigma$ level with the models predictions when convolved with the observational uncertainties. The observed velocities (mostly the $D_n - \sigma$ data), however, exhibit a large tail of high-velocity groups and clusters, to $v_{1D} \sim 3000$ km s$^{-1}$. This high-velocity tail is not consistent with the convolved models (except HDM).

(6) The large uncertainties in the existing measurements of peculiar velocities do not currently allow significant constraints of cosmological models. More accurate peculiar velocity observations of groups and of clusters of galaxies are likely to yield *a lower velocity-tail* than suggested by existing observations. Accurate peculiar velocities of clusters and groups of galaxies will help constrain cosmological models and test whether the initial density field was Gaussian or non-Gaussian.

In summary, we conclude that groups and clusters of galaxies provide robust and efficient tracers of the large-scale peculiar velocity distribution. The current data are consistent at $\sim 3\sigma$ level with the cluster velocity distribution expected from the four models studied. In particular, the observed velocity distribution is consistent with a COBE-normalized low-density CDM-type model, which best fits other large-scale structure observations. It is also consistent with an initial density field that is Gaussian.

## ACKNOWLEDGMENTS

It is a pleasure to acknowledge NCSA for allowing us to use their Convex-3880 supercomputer, on which our computations were performed. We gratefully thank J. Goodman, J.R. Gott, J.P. Ostriker, D.N. Spergel and R. Wijers for useful discussions. This work is supported by NASA grant NAGW-2448, and NSF grants AST90-20506, AST91-08103, AST93-15368 and ASC93-18185.

TABLE 1

MODEL PARAMETERS

| MODEL | $\Omega$ | $\Omega_\Lambda$ | $h$ | $\sigma_8$ |
|---|---|---|---|---|
| CDM | 1.0 | 0.0 | 0.50 | 1.05 |
| CDM | 0.3 | 0.7 | 0.67 | 0.67 |
| HDM | 1.0 | 0.0 | 0.50 | 0.86 |
| PBI | 0.3 | 0.7 | 0.50 | 1.02 |

TABLE 2

3-D RMS PECULIAR VELOCITY OF CLUSTERS, GROUPS AND MATTER
$< v_{3\mathrm{D}}^2 >^{1/2}$

| Model | Rich clusters | Groups | Mass | Smoothed mass* | Linear theory* |
|---|---|---|---|---|---|
| CDM $\Omega = 1.0$ | 762 km s$^{-1}$ | 847 km s$^{-1}$ | 1060 km s$^{-1}$ | 745 km s$^{-1}$ | 784 km s$^{-1}$ |
| CDM $\Omega = 0.3$ | 440 km s$^{-1}$ | 465 km s$^{-1}$ | 483 km s$^{-1}$ | 410 km s$^{-1}$ | 388 km s$^{-1}$ |
| HDM $\Omega = 1.0$ | 815 km s$^{-1}$ | 1060 km s$^{-1}$ | 1022 km s$^{-1}$ | 808 km s$^{-1}$ | 859 km s$^{-1}$ |
| PBI $\Omega = 0.3$ | 712 km s$^{-1}$ | 715 km s$^{-1}$ | 771 km s$^{-1}$ | 648 km s$^{-1}$ | 660 km s$^{-1}$ |

* Smoothed with a Gaussian of radius $R_G = 3h^{-1}$ Mpc.

TABLE 3

1-D RMS PECULIAR VELOCITY OF GROUPS: MODELS VERSUS DATA

$< v_{\rm 1D}^2 >^{1/2}$

| model | model groups (unconvolved)* | | model groups (convolved)* | | observed groups |
|---|---|---|---|---|---|
| CDM $\Omega = 1.0$ | 489 km $s^{-1}$ | Total | 700 km $s^{-1}$ | Total | $725 \pm 50$ km $s^{-1}$ |
| | | TF | 650 km $s^{-1}$ | TF | $607 \pm 64$ km $s^{-1}$ |
| CDM $\Omega = 0.3$ | 268 km $s^{-1}$ | Total | 544 km $s^{-1}$ | Total | $725 \pm 50$ km $s^{-1}$ |
| | | TF | 477 km $s^{-1}$ | TF | $607 \pm 64$ km $s^{-1}$ |
| HDM $\Omega = 1.0$ | 614 km $s^{-1}$ | Total | 812 km $s^{-1}$ | Total | $725 \pm 50$ km $s^{-1}$ |
| | | TF | 769 km $s^{-1}$ | TF | $607 \pm 64$ km $s^{-1}$ |
| PBI $\Omega = 0.3$ | 413 km $s^{-1}$ | Total | 618 km $s^{-1}$ | Total | $725 \pm 50$ km $s^{-1}$ |
| | | TF | 565 km $s^{-1}$ | TF | $607 \pm 64$ km $s^{-1}$ |

* Convolved (or unconvolved) with the observational uncertainties. The total sample and TF sample are listed separately, as indicated.